\documentclass[12pt,twocolumn,tighten]{aastex62}
\usepackage{amsmath,amstext,amssymb}
\usepackage[T1]{fontenc}
\usepackage{apjfonts}
\usepackage[figure,figure*]{hypcap}
\usepackage{graphics,graphicx}
\usepackage{hyperref}
\usepackage{natbib}
\usepackage[caption=false]{subfig} 
\usepackage{enumitem} 
\usepackage{epigraph}

\newcommand{\ptfo}{PTFO$\,$8-8695}
\newcommand{\ptfob}{PTFO$\,$8-8695b}

\received{May 15, 2020}
\revised{\today}
\accepted{---}
\submitjournal{AAS journals.}
\shorttitle{Two Stars, Two Signals, No Planet}

\begin{document}

\defcitealias{bouma_wasp4b_2019}{B19}

\title{PTFO 8-8695: Two Stars, Two Signals, No Planet}

\correspondingauthor{L.\,G.\,Bouma}
\email{luke@astro.princeton.edu}

%
%
\author[0000-0002-0514-5538]{L.\,G.\,Bouma}
\affiliation{Department of Astrophysical Sciences, Princeton
University, 4 Ivy Lane, Princeton, NJ 08540, USA}
\author[0000-0002-4265-047X]{J.\,N.\,Winn}
\affiliation{Department of Astrophysical Sciences, Princeton
University, 4 Ivy Lane, Princeton, NJ 08540, USA}
%
%
%
\author{G. R. Ricker} 
\affiliation{Department of Physics and Kavli Institute for
Astrophysics and Space Research, Massachusetts Institute of
Technology, Cambridge, MA 02139, USA}
%
\author[0000-0001-6763-6562]{R. Vanderspek} 
\affiliation{Department of Physics and Kavli Institute for
Astrophysics and Space Research, Massachusetts Institute of
Technology, Cambridge, MA 02139, USA}
%
\author[0000-0001-9911-7388]{D. W.~Latham} 
\affiliation{Center for Astrophysics \textbar \ Harvard \&
Smithsonian, 60 Garden St, Cambridge, MA 02138, USA}
%
\author[0000-0002-6892-6948]{S.~Seager} 
\affiliation{Department of Physics and Kavli Institute for
Astrophysics and Space Research, Massachusetts Institute of
Technology, Cambridge, MA 02139, USA}
\affiliation{Department of Earth, Atmospheric and Planetary Sciences,
Massachusetts Institute of Technology, Cambridge, MA 02139, USA}
\affiliation{Department of Aeronautics and Astronautics, MIT, 77
Massachusetts Avenue, Cambridge, MA 02139, USA}
%
\author[0000-0002-4715-9460]{J. M.~Jenkins} 
\affiliation{NASA Ames Research Center, Moffett Field, CA 94035, USA}
%

\author[0000-0001-7139-2724]{T. Barclay} 
\affiliation{NASA Goddard Space Flight Center, 8800 Greenbelt Road,
Greenbelt, MD 20771, USA}
\affiliation{University of Maryland, Baltimore County, 1000 Hilltop
Circle, Baltimore, MD 21250, USA}

\author[0000-0001-6588-9574]{K. A. Collins} 
\affiliation{Center for Astrophysics \textbar \ Harvard \&
Smithsonian, 60 Garden St, Cambridge, MA 02138, USA}

\author{J. P. Doty} 
\affiliation{Noqsi Aerospace Ltd., 15 Blanchard Avenue, Billerica, MA,
01821, USA}

\author[0000-0002-2457-272X]{D.~R.~Louie} 
\affiliation{Department of Astronomy, University of Maryland, College
Park, MD 20742, USA}

\author[0000-0002-8964-8377]{S. N. Quinn} 
\affiliation{Center for Astrophysics \textbar \ Harvard \&
Smithsonian, 60 Garden St, Cambridge, MA 02138, USA}

\author[0000-0003-4724-745X]{M. E. Rose} 
\affiliation{NASA Ames Research Center, Moffett Field, CA 94035, USA}

\author[0000-0002-6148-7903]{J. C. Smith} 
\affiliation{NASA Ames Research Center, Moffett Field, CA 94035, USA}
\affiliation{SETI Institute, Mountain View, CA 94043, USA}

\author{J. Villase\~nor} 
\affiliation{Department of Physics and Kavli Institute for
Astrophysics and Space Research, Massachusetts Institute of
Technology, Cambridge, MA 02139, USA}

\author[0000-0002-5402-9613]{B. Wohler} 
\affiliation{NASA Ames Research Center, Moffett Field, CA 94035, USA}
\affiliation{SETI Institute, Mountain View, CA 94043, USA}

\begin{abstract}
  PTFO$\,$8-8695 (CVSO\,30) is a star in the 7--10 million year old
  Orion-OB1a cluster that shows brightness dips that resemble
  planetary transits.  Although strong evidence against the planet
  hypothesis has been presented, the possibility remains debated in
  the literature.  To obtain further clues, we inspected data from the
  NASA {\it Transiting Exoplanet Survey Satellite} (TESS) and the ESA
  Gaia mission.  The Gaia data suggest that PTFO$\,$8-8695 is a
  binary: the photometric data show it to be overluminous with respect
  to members of its kinematic group, and the astrometric data are
  inconsistent with a single star.  The TESS light curve shows two
  different photometric periods. The variability is dominated by a
  sinusoidal signal with a period of 11.98$\,$hr, presumably caused by
  stellar rotation.  Also present is a 10.76$\,$hr signal consisting
  of a not-quite sinusoid interrupted by hour-long dips, the type of
  signal previously interpreted as planetary transits.  The phase of
  the dips is nearly 180$^\circ$ away from the phase of the originally
  reported dips. As noted previously, this makes them difficult to
  explain as planetary transits.  Instead, we believe that
  PTFO$\,$8-8695 is a pair of young and rapidly rotating M dwarfs, one
  of which shows the same ``transient-dipper'' behavior that has been
  seen in at least 5 other cases.  The origin of these transient dips
  is still unknown but likely involves circumstellar material.
\end{abstract}



\section{Introduction}

We wish \ptfob\ were a planet. It would be quite exceptional. It would
be the youngest known hot Jupiter \citep{van_eyken_ptf_2012}, orbiting
a T Tauri star in the Orion-OB1a cluster.  It would have the shortest
orbital period of any hot Jupiter.  With such a short period, it would
probably be filling its Roche lobe, and actively losing mass to its
host star.  Not only that, but the rapidly-rotating host star is
probably oblate enough to torque the planet's orbit into and out of
the transiting configuration on a timescale of years
\citep{barnes_measurement_2013,ciardi_followup_2015,kamiaka_revisiting_2015}. 

Another first would be the direct detection of H$\alpha$ emission from
the planet itself \citep{johnskrull_h_2016}.  In addition to the
chromospheric H$\alpha$ emisson, it seems that there is an additional
H$\alpha$ emission with radial velocity variations in phase with the
planetary orbit.  The average velocity width of the excess H$\alpha$
emission is 87$\,$km$\,$s$^{-1}$, and its equivalent width is 70-80\%
that of the stellar chromosphere \citep{johnskrull_h_2016}.  The
proposed explanation is that the emission is from hot material flowing
away from the planet \citep{johnskrull_h_2016}.

However, the observed signals have some peculiarities that make the
planet seem even more unusual, to the point that they cast into doubt
the premise that \ptfob\ is real.  First, the transit-like brightness
dips are about three times deeper in optical bandpasses ({\it e.g.,}
$g$-band) than in the near-infrared ({\it e.g.}, $z$-band)
\citep{onitsuka_multicolor_2017,tanimoto_evidence_2020}.  An ordinary
atmosphere expected for a Jovian planet would not lead to such a
strong color-dependence of the transits.  Second, the planet does not
seem to emit as much infrared radiation as would be expected for such
a hot Jovian planet \citep{yu_tests_2015}.  Third, despite measurement
attempts by multiple investigators, \ptfob\ does not seem to show the
Rossiter effect at the amplitude expected given the rapid stellar
rotation and large planet size
\citep{yu_tests_2015,ciardi_followup_2015}.  Fourth, the phase of the
dips within the overall period of photometric variability has changed
drastically over the years since their initial discovery.  To counter
these objections, it has been proposed that the planet may be much
smaller than Jupiter and that the dips are produced by dust clouds
emitted from the planet \citep{tanimoto_evidence_2020}. 

A separate issue is that the brightness dips change shape over many
orbital cycles. This was initially explained by
\cite{barnes_measurement_2013} as the natural effects of gravity
darkening.  However, \cite{howarth_reappraisal_2016} argued that the
necessary amplitude of gravity darkening is too large to be realistic,
given the spectroscopically-determined rotation velocity.
Additionally, as the gravity-darkened star precessed about its
rotation axis, it would show photometric variability that has not been
observed.

While the planetary interpretation clearly faces challenges, there is
no completely satisfactory alternate explanation.  Low-latitude
starspots, hot or cold, would struggle to produce photometric features
as short as some of the observed dips.  High-latitude accretion
hotspots might produce the observed H$\alpha$ variability, but require
fine-tuning to produce dips of the appropriate duration.  Transits by
dust clumps or other dusty features are questionable because \ptfo\
does not have a detectable infrared (IR) excess associated with the
presence of warm dust \citep[{\it e.g.},][Figure~18]{yu_tests_2015}.
In addition, the sublimation times for dust grains of plausible
composition are quite short \citep{zhan_complex_2019}. 

A relevant fact is that between 0.1\% and 1\% of rapidly rotating
low-mass stars in $\mathcal{O}$(10)$\,$Myr old associations show
short-duration dips as part of their overall periodic variability
\citep{rebull_usco_2018}.  The dips can persist over months, but their
depths often vary, and sometimes change immediately after stellar
flares.  The explanation proposed by \citet{stauffer_orbiting_2017}
and \citet{david_transient_2017} to explain this novel class of
variable stars is that a circumstellar cloud of gas is orbiting near
the co-rotation radius.  To this point, though, it has not been clear
if this explanation applies to \ptfo, because the determination of the
stellar rotation period has been somewhat ambiguous
\citep{van_eyken_ptf_2012,koen_multicolour_2015,raetz_yeti_2016}.

We begin in Section~\ref{sec:observations} by describing newly
available observations from TESS \citep{ricker_transiting_2015} and
Gaia \citep{gaia_collaboration_gaia_2018}.  The TESS light curve shows
two different periodic signals, which we analyze in
Section~\ref{sec:tess}.  The Gaia data, analyzed in
Section~\ref{sec:gaia}, show that \ptfo\ is too bright to be a single
star and also suggest it is an astrometric binary.  We discuss the
pieces of the puzzle in Section~\ref{sec:discussion}, and summarize
the situation in Section~\ref{sec:conclusions}.  In a postscript, we
comment on a recent study by \citet{koen_2020} which reached similar
conclusions.

\section{The Data}
\label{sec:observations}

\begin{figure*}[t!]
	\begin{center}
		\leavevmode
		\includegraphics[width=1\textwidth]{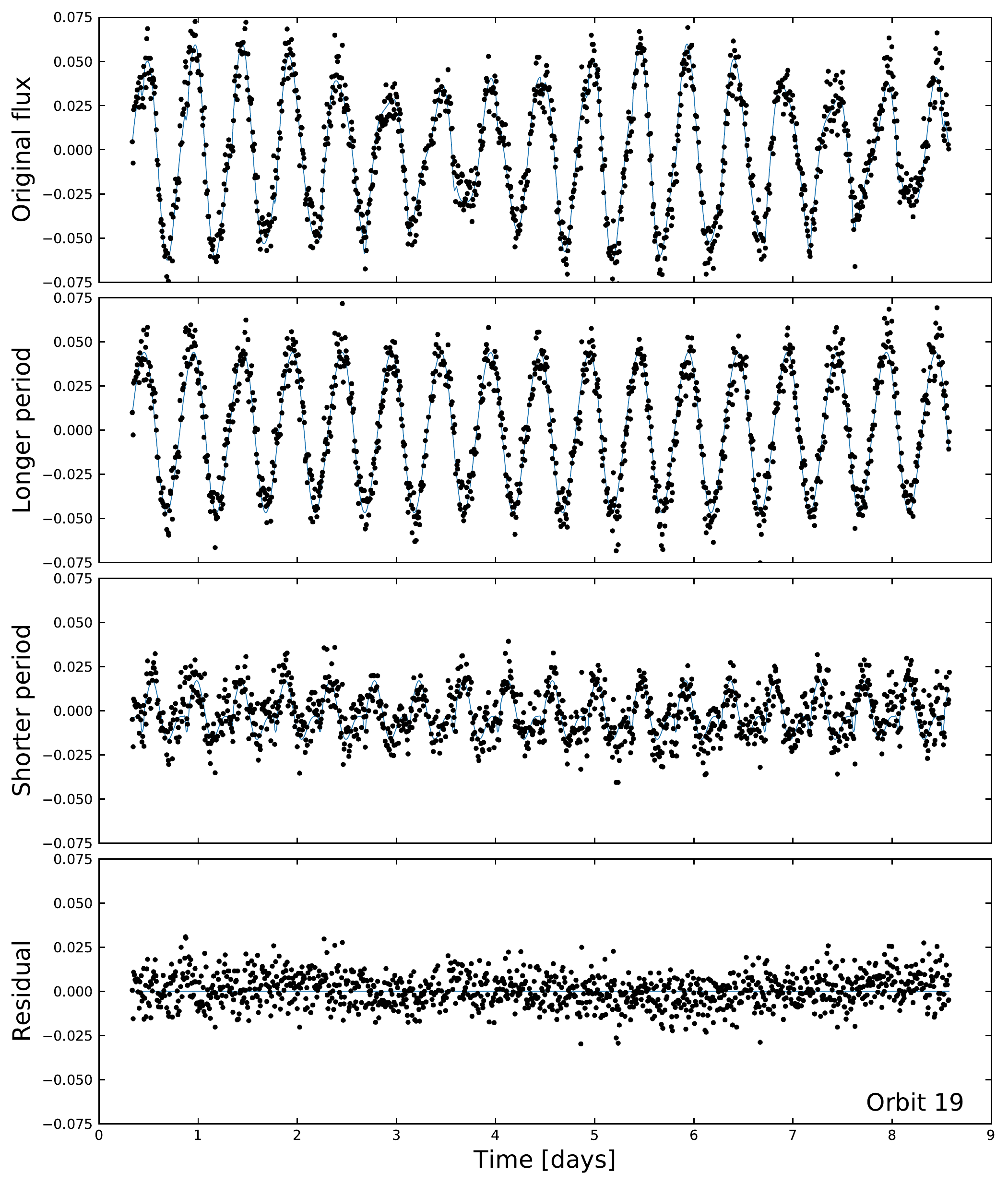}
	\end{center}
	\vspace{-0.7cm}
	\caption{
    {\bf TESS light curve of \ptfo\ (Sector 6, Orbit 19).} {\it Top}:
    The original (\texttt{PDCSAP} median-subtracted) relative flux.
    The beat period of 4.48 days is visible by eye.  The blue curve is
    a model including 2 harmonics at the longer period $P_{\rm \ell}$,
    plus 3 harmonics and a transit at the shorter period $P_{\rm s}$.
    {\it Upper middle}: Longer-period signal, equal to the original
    signal minus the shorter-period signal.  {\it Lower middle}:
    Shorter-period signal, equal to the original signal minus the
    longer-period signal.  {\it Bottom}: residual relative flux.  The
    data are binned from 2 to 10 minute cadence for convenience in
    plotting and fitting.
		\label{fig:splitsignal}
	}
\end{figure*}

\begin{figure*}[hbtp]
	\begin{center}
		\leavevmode
		\includegraphics[width=1\textwidth]{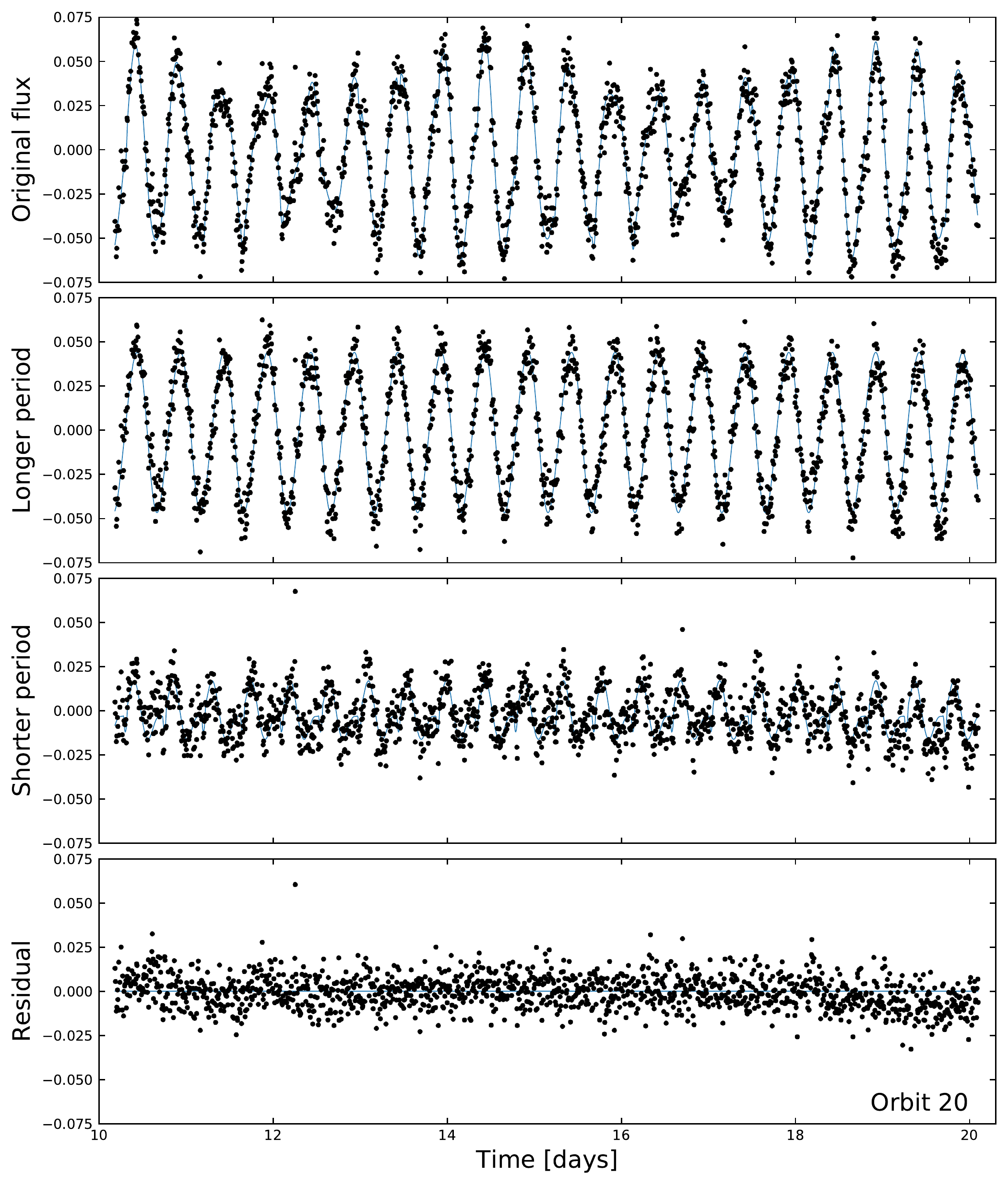}
	\end{center}
	\vspace{-0.7cm}
  \caption{ {\bf TESS light curve of \ptfo\ (Sector 6, Orbit 20).}
  Same format as Figure~\ref{fig:splitsignal}.
  \label{fig:splitsignalii}
	}
\end{figure*}

\subsection{TESS Observations}

\ptfo\ (also known as CVSO\,30; \citealt{briceno_cida_2005}) was
observed by TESS with Camera 1, CCD 1, from December 15, 2018 until
January 6, 2019, during the sixth sector of science operations
\citep{ricker_transiting_2015}.  The star is designated TIC\,264461976
in the TESS Input Catalog \citep{stassun_TIC_2018,stassun_TIC8_2019}.
The pixel data for an $11\times11$ array surrounding \ptfo\ were
averaged into 2-minute stacks by the onboard computer.  Each
2048$\times$2048 image from the CCD was also averaged into 30-minute
stacks, and saved as a ``full frame image'' (FFI).

The 2-minute stacks for \ptfo\ were reduced to light curves by the
Science Processing Operations Center (SPOC) at the NASA Ames Research
Center~\citep{jenkins_tess_2016}.  We mainly used the Presearch Data
Conditioning Simple Aperture Photometry (PDCSAP) light curve.  The PDC
light curve is based on pixels chosen to maximize the SNR of the total
flux of the target \citep{bryson_2020_target_aperture}.
Non-astrophysical variability was removed by fitting out trends common
to many stars \citep{smith_kepler_2012,stumpe_multiscale_2014}.

As an independent check on the 2-minute SPOC light curve, we examined
the light curve based upon 30-minute image stacks which was produced
as part of the Cluster Difference Imaging Photometric Survey (CDIPS;
\citealt{bouma_cluster_2019}).  Our CDIPS light curve of choice used a
circular aperture with radius 1 pixel.

To clean the data, we removed all points with non-zero quality flags,
which indicate known problems \citep[{\it
e.g.},][]{tess_data_product_description_2018}.  We also masked out the
data from the first and last 6 hours of each orbit, since there are
often systematic effects in the photometry during those times.  Both
the CDIPS and PDC light curves showed a discontinuous jump in the last
few days of orbit 20, which seemed likely to be an instrumental
systematic effect, and led us to mask out the data with timestamps
ranging from BJD\,2458488.3 until the end of the orbit.  The PDC light
curve initially had 15{,}678 points.  The quality-flag cut removed 854
points; masking the orbit edges removed an additional 716 points; and
removing the data from the final few days of orbit 20 removed an
additional 1{,}079 points.  After cleaning, 83\% of the initial flux
measurements remained.

We normalized the light curve by dividing out the median flux, and
then opted to subtract 1{.}0 to set the median value to zero, which
simplified subsequent interpretation.  Many of these and subsequent
processing steps were performed using
\texttt{astrobase}~\citep{bhatti_astrobase_2018}.

\subsection{Gaia Observations}

\subsubsection{Astrometric measurements}

Between July 25, 2014 and May 23, 2016, Gaia measured about 300
billion centroid positions of 1{.}6 billion stars
\citep{gaia_collaboration_gaia_2016,lindegren_gaiasoln_2018,gaia_collaboration_gaia_2018}.
For the Gaia second data release (DR2), these CCD observations were
used to determine positions, proper motions, and parallaxes of the
brighest 1{.}3 billion stars \citep{lindegren_gaiasoln_2018}.  For
\ptfo, there were 121 ``good'' observations, {\it i.e.}, observations
that were not strongly down-weighted in the astrometric solution.
\ptfo\ was assigned the Gaia DR2 identifier 3222255959210123904.  Its
brightness was measured using selected bands ($G$, $Rp$, and $Bp$) of
the Gaia Radial Velocity Spectrometer
\citep{cropper_gaia_2018,evans_gaia_2018}.  We accessed the pipeline
parameters for \ptfo\ using the Gaia
archive\footnote{\url{gea.esac.esa.int/archive/}}.

The majority of Gaia's derived parameters for \ptfo\ agree with
expectations based on previous studies
\citep{briceno_cida_2005,van_eyken_ptf_2012}.  The main novelty is
that Gaia DR2 reported a 10.3$\sigma$ ``astrometric excess'',
indicating that the residuals to the best-fitting astrometric model
were larger than expected based on the statistical uncertainties.  We
comment on the significance and interpretation of this excess in
Section~\ref{sec:gaia}.

\subsubsection{Hierarchical Cluster Membership}
\label{subsec:hierarchical}

Gaia also provided astrometric parameters for tens of thousands of
young stars in the Orion complex.  Stellar populations in giant
molecular cloud complexes are not monolithic; substructured groups are
the norm \citep{briceno_lowmassOB_2007}.  The Orion molecular cloud
complex in particular has numerous subgroups, with ages ranging from
0.5 to 15$\,$Myr. See, for instance,
\citet{briceno_cida_2005,jeffries_kinematic_2006,briceno_25_2007,kounkel_apogee2_2018}
and \citet{briceno_cidaII_2019}.

\ptfo\ was initially designated CVSO\,30 and identified as a member of
the Orion$\,$OB1a sub-association by \citet{briceno_cida_2005}, based
on photometry and spectroscopy.  Later work by \citet{briceno_25_2007}
clarified that \ptfo\ is in a kinematically distinct subgroup of
Orion~OB1a, named the ``25$\,$Ori'' group after its brightest member.
They reported that the 25$\,$Ori group has an isochrone age of
7--10$\,$Myr, and a smaller fraction of stars with disks than younger
nearby sub-associations \citep{hernandez_spitzer_ob1_2007}.

With the Gaia astrometry, it has become clear that 25$\,$Ori itself
has distinct subgroups
\citep{kounkel_apogee2_2018,briceno_cidaII_2019}.  In describing the
cluster membership of \ptfo, we follow the notation and results of
\citet{kounkel_apogee2_2018}.  These authors combined astrometric data
from Gaia DR2 with near-infrared spectra from APOGEE-2
\citep{gunn_sdss_2006,majewski_apache_2017,blanton_sloan_2017,zasowski_target_2017,cottle_apogee2_2018}.
They performed a hierarchical clustering on the six-dimensional
position and velocity information to identify subgroups within the
Orion complex.  From smallest to largest, \ptfo\ was identified as a
member of the following hierarchical subgroups:
\begin{equation}
  {\rm 25\,Ori\text{-}1}
  \subset {\rm 25\,Ori}
  \subset {\rm Orion\ OB1a}
  \subset {\rm Orion\ D},
\end{equation}
where `$\subset$' means `is a proper subset of'.  25$\,$Ori-1 is the
largest subgroup of 25$\,$Ori, with 149 identified members.  The mean
age of the 25$\,$Ori-1 subgroup, determined by fitting isochrones to
group members with APOGEE effective temperatures and Gaia parallaxes,
was determined to be $8.5\pm1.2\,{\rm Myr}$ \citep[see][Section
2.3]{kounkel_apogee2_2018}.  \citet{kounkel_apogee2_2018} also
identified seven smaller groups in the Orion complex near the Be star
25$\,$Ori. These groups received sequential identifiers, {\it e.g.},
25$\,$Ori-2 (${\rm Age} =12.9\pm2.8\,$Myr; see also
\citealt{briceno_cidaII_2019}).

These details concerning the group membership for one object may seem
excessive to those accustomed to the simple distinction between
``young cluster members'' and ``old field stars''.  Although all
members of the Orion complex are indeed young relative to the field,
these details are essential for assessing the photometric evidence for
the binarity of \ptfo, because of the degeneracy between stellar
luminosity and age for pre-main-sequence stars.  Having a clean sample
of reference stars that are tightly associated with \ptfo\ --- both
spatially and kinematically --- minimizes contamination not only from
field stars, but also from older and younger members of the Orion
complex.

\section{TESS Analysis}
\label{sec:tess}

\begin{figure*}[t]
	\begin{center}
		\leavevmode
		\includegraphics[width=.93\textwidth]{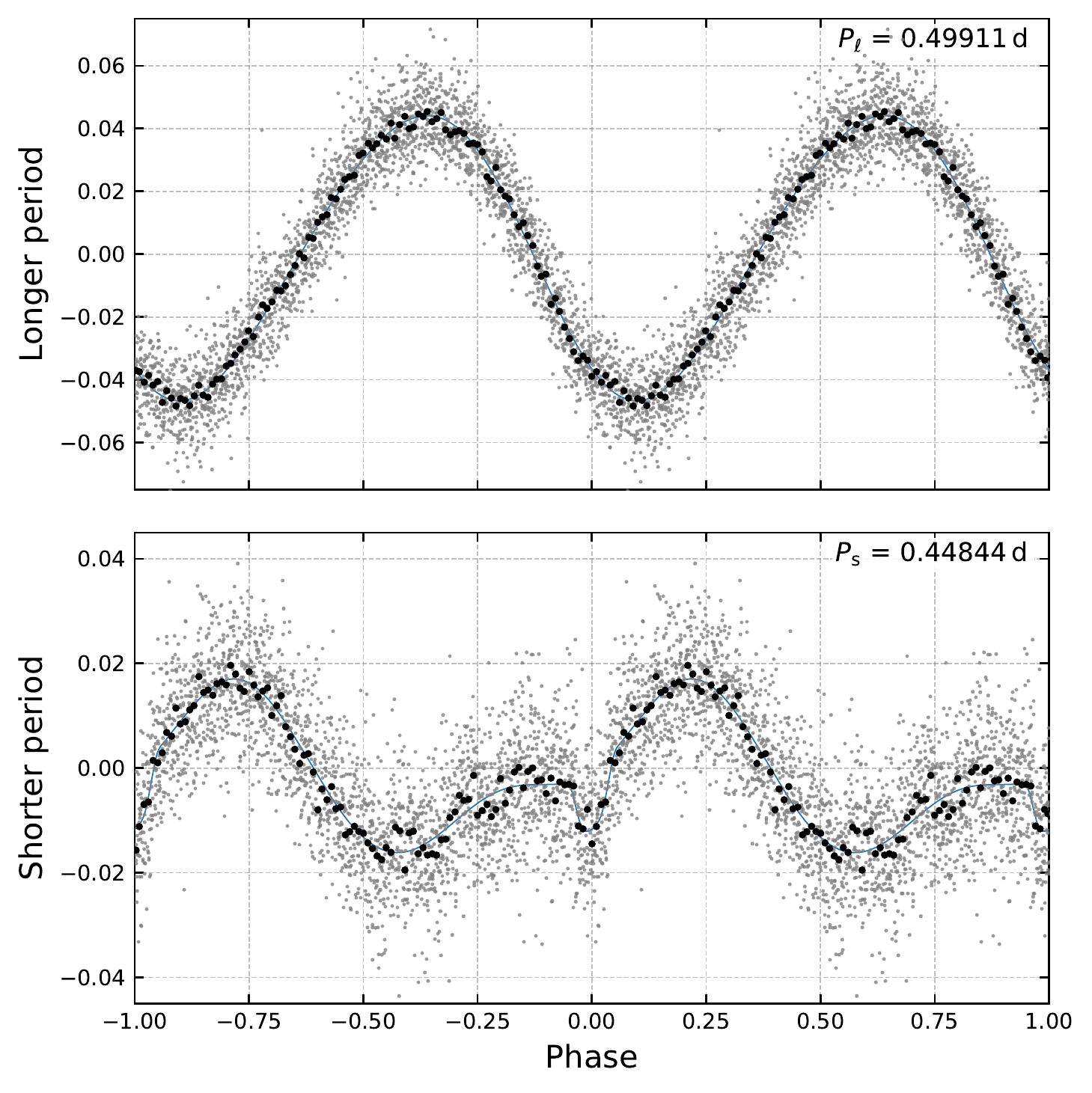}
	\end{center}
	\vspace{-0.7cm}
	\caption{ {\bf Phase-folded longer and shorter-period signals.}
    {\it Top}: The longer-period signal.  {\it Bottom}: The
    shorter-period signal. The phase is defined such that the dip
    occurs at zero phase.  Gray points are the \texttt{PDCSAP} data
    binned to 10-minute cadence.  Black points are binned to 100
    points per period.  The model (blue line) includes 2 harmonics at
    the longer period, plus 3 harmonics and a transit at the shorter
    period.
		\label{fig:phasefold}
	}
\end{figure*}

\subsection{Inspection}

Our initial inspection of the TESS light curve, in both its 2-minute
PDCSAP and 30-minute FFI forms, showed a strong sinusoidal beat signal
(Figures~\ref{fig:splitsignal} and~\ref{fig:splitsignalii}, top
panel). As a precursor to more detailed analysis, we calculated
generalized Lomb-Scargle periodograms using \texttt{astrobase}
\citep{lomb_1976,scargle_studies_1982,vanderplas_periodograms_2015,bhatti_astrobase_2018}.
The tallest peak occurs at 0.499\,d (11.98$\,$hr) and a second strong
peak occurs at 0.448\,d (10.76$\,$hr). We will refer to these two
periods as the ``longer period'' $P_{\rm \ell}$ and the ``shorter
period'' $P_{\rm s}$.  Lower-power harmonics of both signals are also
present.

The peak-to-peak maximum amplitude of the light curve, when the two
signals interfere constructively, is about 14\%.  During the times of
destructive interference, the peak-to-peak amplitude is about 6\%.
Assuming the signals are mainly sinusoidal, simple algebra tells us
that the peak-to-peak amplitudes should be about 10\% for the
longer-period signal, and 4\% for the shorter-period signal.  To view
the phase-folded light curves of the longer-period signal, we
subtracted the best-fitting sinusoid with a period equal to $P_{\rm
s}$. The resulting light curve appears smooth and nearly sinusoidal.
But after subtracting the best-fitting sinusoid with a period equal to
$P_{\ell}$, visual inspection of the phase-folded light curve revealed
substructure resembling the ``dips'' seen in previous observations. In
particular, there was a $\approx$1\% dip lasting about an hour.  These
initial impressions turned out to be consistent with the results of
our more complicated analysis, described below.

\subsection{Light Curve Model}

We fitted a model to the light curve consisting of a linear combination
of Fourier modes with periods $P_{\rm s}$ and $P_{\rm \ell}$, as well
as a number of harmonics chosen as described below. To try accounting
for the dips, we also added an analytic transit model with period
$P_{\rm s}$.  Symbolically, the total flux $f$ is given as
\begin{equation}
  f = f_{\rm s} + f_{\rm \ell}
  = f_{\rm transit,s} + f_{\rm Fourier,s} + f_{\rm Fourier,\ell},
\end{equation}
where $f_{\rm s}$ is the flux at the shorter period, and $f_{\rm
\ell}$ is the flux at the longer period.  Writing out the Fourier
terms explicitly,
\begin{align}
  f = &f_{\rm transit,s} + \sum_{n=1}^{N_{\rm s}} A_n \sin(n\omega_{\rm s}t)
  + \sum_{n=1}^{N_{\rm s}} B_n \cos(n\omega_{\rm s}t)\\
  &+ \sum_{m=1}^{N_{\rm \ell}} A_m \sin(m[\omega_{\rm \ell}t+\phi_{\rm \ell}])
  + \sum_{m=1}^{N_{\rm \ell}} B_m \cos(m[\omega_{\rm \ell}t+\phi_{\rm \ell}]), \nonumber
\end{align}
where $N_{\rm s}$ and $N_{\rm \ell}$ are the total number of modes at
the shorter and longer periods, respectively, $A_i$ and $B_i$ are the
amplitudes of each mode (which can be positive or negative), and
$\omega_\ell$ and $\omega_{\rm s}$ are the angular frequencies of the
longer-period and shorter-period signals. By not including a phase
parameter in the shorter-period model, we have implicitly defined the
zero point of the phase scale. The relative phase of the longer-period
model is specified by the phase parameter $\phi_\ell$.  Since we did
not know in advance how many harmonics would be appropriate to include
in the model, we considered a number of different choices for $N_{\rm
s}$ and $N_{\rm \ell}$, and used the Bayesian information criterion to
select the final model (Table~1).

The free parameters are as follows.  The transit model parameters are
the impact parameter, the planet-to-star radius ratio, two quadratic
limb darkening parameters, the planet's orbital period (set equal to
$P_{\rm s}$) the time of a particular transit, and the mean flux.  We
sampled the stellar radius and mass from prior probability
distributions, implicitly defining the stellar density which (together
with the orbital period) sets the transit timescale.  There are also
the parameters defining the Fourier modes.  As an example, one
possible model consists of a transit, $N_{\rm s}=2$ sines and cosines
at the shorter period, plus $N_{\rm \ell}=1$ sine and cosine at the
longer period.  There are $2N_{\rm s}=4$ additional Fourier amplitudes
at the shorter period, plus $2N_{\rm \ell}=2$ Fourier amplitudes at
the longer period, as well as $P_\ell$ itself and the relative phase
$\phi_\ell$.  The total number of parameters is 17 for this case.

We implemented and fitted the models using \texttt{PyMC3}, which is
built on \texttt{theano}
\citep{salvatier_2016_PyMC3,exoplanet:theano}.  For the Fourier terms,
we used the default math operators.  For the exoplanet transit, we
used the model and derivatives implemented in the \texttt{exoplanet}
code \citep{exoplanet:exoplanet}.  Our priors are listed in Table~2.
To speed up the fitting process, we averaged the 2-minute light curve
into 10-minute samples.  We correspondingly scaled down the
uncertainties in the flux measurements by a factor of $\sqrt{5}$.
Before sampling, we initialized each model with the parameters of the
maximum {\it a posteriori} (MAP) model.  We then assumed a Gaussian
likelihood, and sampled using \texttt{PyMC3}'s gradient-based
No-U-Turn Sampler \citep{hoffman_no-u-turn_2014}, and used $\hat{R}$
as our convergence diagnostic \citep{gelman_inference_1992}.  We
tested our ability to successfully recover injected parameters using
synthetic data before fitting the \ptfo\ light curves.

\subsection{Fitting Results}

We considered nine models, with the number of modes per frequency
($N_{\rm s}$ and $N_{\rm \ell}$) ranging from one to three.  To select
our preferred model, we used the Bayesian information criterion
(Table~1).  The model with the lowest BIC had three modes at the
shorter 10.76$\,$hr period, and two modes at the longer 11.98$\,$hr
period.  The other models had BIC values that implied significantly
less support \citep{burnham_multimodel_2016}.  All nine models have
reduced $\chi^2$ values ranging between 1.21 and 1.68, which suggests
a plausible though imperfect agreement between the data and the model
to within the formal uncertainties.  Table~2 gives the best-fitting
parameters for the preferred model, which has the lowest BIC value.

To explore where each model succeeded and failed, we split the
original signal into its respective components
(Figures~\ref{fig:splitsignal} and~\ref{fig:splitsignalii}).  We also
examined the phase-folded signals (Figure~\ref{fig:phasefold}).  

In every model, the 11.98$\,$hr variability is a simple sinusoid with
peak-to-peak amplitude $\approx$10\%.  The 10.76$\,$hr variability is
always more complex.  The overall impression is of a distorted
sinusoidal function, with a peak-to-peak amplitude of about 4\%.  The
asymmetric sinusoid rises to a maximum near phase 0.25, and reaches
minimum brightness between phases $-0.5$ and $-0.25$.  Between phases
$-0.5$ and $0.0$ there appears to be complex shorter-timescale
variability, ending with a ``dip'' of depth $\approx$1.2\%, lasting
$\approx$0.75 hours.  The fact that our preferred model has three
rather than two ``short period'' harmonics is linked to the degree of
curvature required between phases $-0.5$ and $-0.05$: the analogous
$(N_{\rm \ell},N_{\rm s})=(2,2)$ model prefers a longer transit
duration, but does not fit the out-of-transit curvature as well,
particularly immediately before ingress.

The periodogram of the residuals between the data and the preferred
model shows a barely significant and poorly-resolved peak at
$\approx$8 days, consistent with the visual impression of some slower
trends in the bottom rows of Figures~\ref{fig:splitsignal}
and~\ref{fig:splitsignalii}.

\section{Tests for Binarity}
\label{sec:gaia}

\subsection{Visual Binarity}
\label{subsec:blend}

\begin{figure}[t]
	\begin{center}
		\leavevmode
		\includegraphics[width=0.45\textwidth]{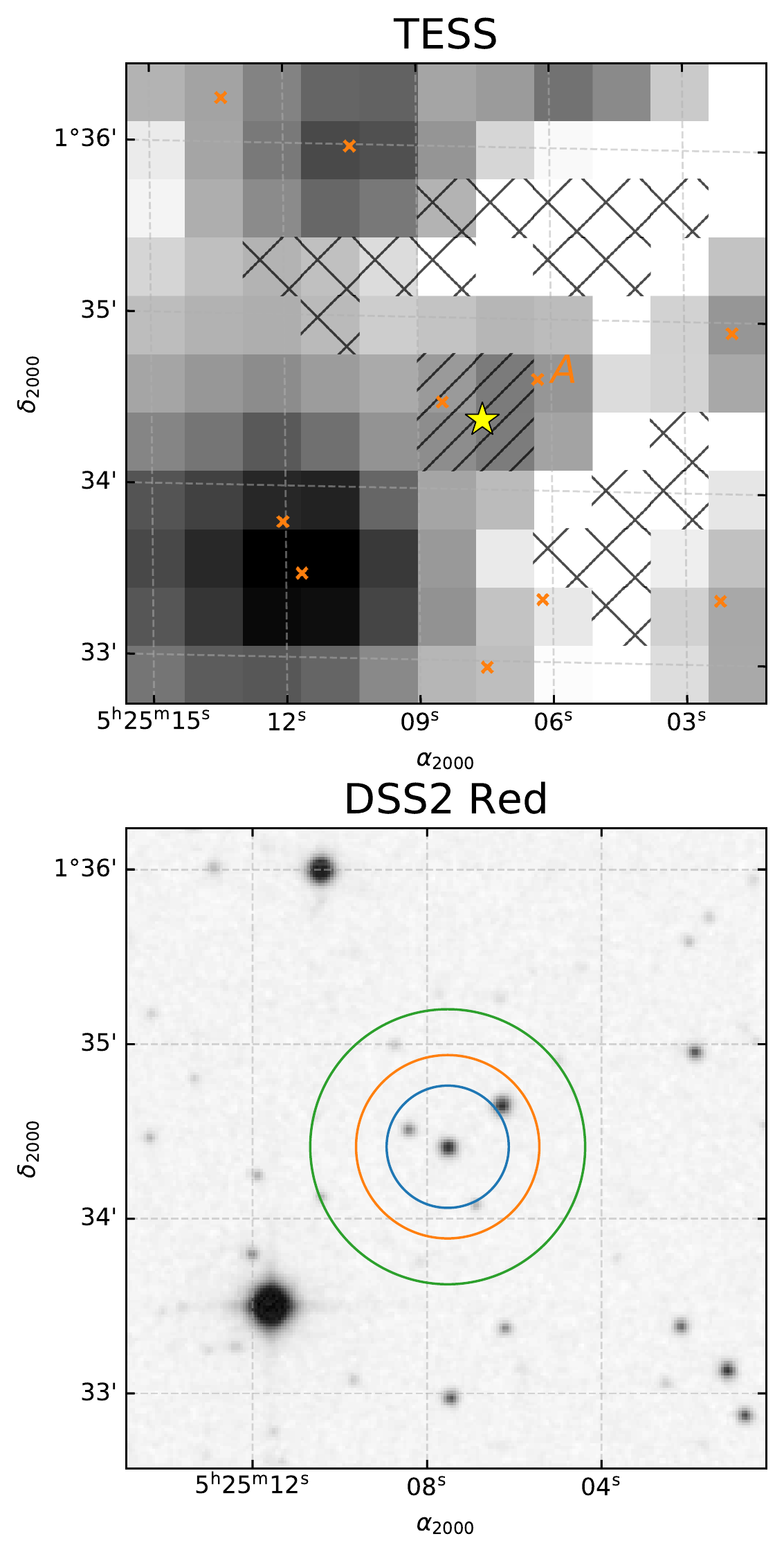}
	\end{center}
	\vspace{-0.7cm}
	\caption{ {\bf Scene used for blend analysis.}
    {\it Top:} Mean TESS image of \ptfo\ over Sector~6, with a
    logarithmic grayscale. The yellow star is the position of \ptfo.
    Orange crosses are neighboring stars with $T<17$. The \texttt{X}
    and \texttt{/} hatches show the apertures used to measure the
    background and target star flux, respectively.  {\it Bottom:}
    Digitized Sky Survey $R$-band image of the same field, with a
    linear grayscale. The circles show the apertures of radii 1, 1.5,
    and 2.25 pixels used in our blend analysis. To the northwest of
    \ptfo\ and between the blue and orange circles is ``Star A'', the
    only star bright and close enough to be contributing to the signal
    attributed to \ptfo. However, the pixel-level TESS data showed
    that Star A is not the source of the observed variability (see
    Section~\ref{subsec:blend}).
		\label{fig:scene}
	}
\end{figure}

The portion of the sky subtended by each TESS pixel is about 21'' on a
side. Before making any interpretations, we needed to consider whether
light from neighboring stars could have contributed to the photometric
signal we are attributing to \ptfo. The scene is shown in
Figure~\ref{fig:scene}.  In the upper panels, the pixels used to
measure the background level in the SPOC light curve are indicated
with `\texttt{X}' hatching, and the pixels used in the final light
curve aperture are shown with `\texttt{/}' hatching.

The target star, \ptfo\ (TIC 264461976), has a $T$-band magnitude of
14.0, and its position is shown with a star.  The other (unlabeled)
star inside the target aperture, TIC 264461979, has $T=16.8$ and so
cannot contribute more than about 10\% to the total signal.  The only
other known star that is sufficiently close and bright that its light
might contaminate the signal from the target star is TIC 264461980,
with $T=14.8$.  This star, we we dub ``Star A'', is 23.6'' northwest
of the target. Based on the magnitude difference, Star A could
contribute flux variations as large as 48\% of the flux of the target
star.

The variability of \ptfo\ with a period consistent with $P_{\rm s}$
had already been observed based on images with arcsecond resolution.
Thus, our main concern regarding blending was whether the
longer-period signal with period $P_{\rm \ell}$ originated from \ptfo,
or from Star A. We took two approaches to investigate the source of
the long-period signal.

First, we examined the CDIPS FFI light curves of the target, which are
available on MAST \citep{bouma_cluster_2019}. Three light curves are
available, based on photometric apertures with a radius of 1, 1.5, or
2.5 pixels. The maximal peak-to-peak beat amplitude was the same to
within a percent, regardless of the size of the photometric aperture
that was used to create the light curve.  If Star A were the source of
the long-period variability, we would expect the peak variability
amplitude to be smallest in the 1 pixel aperture, based on the
separation of the sources (Figure~\ref{fig:scene}, bottom).  From this
test alone, it seems unlikely that Star A is the source of the
long-period signal.

Second, we examined the 2-minute light curve of each individual pixel
in the scene, using the interactive tools implemented in
\texttt{lightkurve} \citep{lightkurve_2018}.  If Star A were the
source of the long-period variability, we would expect the pixels
nearest to Star A to show a sinusoidal signal with amplitude exceeding
$10\%$.  The data do not show this pattern.  The data from the pixel
directly below Star A does not show any sinusoidal variability; the
peak-to-peak variability seen in that pixel is $\lesssim 8\%$.  In
contrast, the southeastern-most pixel within the \ptfo\ aperture (the
pixel furthest from Star A that was used in the optimal aperture)
shows the longer-period sinusoidal variability signal with an
amplitude of 14\%.  We conclude that within the resolution of the Gaia
DR2 source catalog, the $P_{\rm s}$ and $P_{\rm \ell}$ signals
originate from \ptfo.  Based on the work of
\citet{ziegler_measuring_2018}, we can surmise that stellar companions
with separations wider than $\approx$1'' (349~AU) and within $\Delta G
\approx 3$ magnitudes of \ptfo\ would have likely been detected
through this approach. 

Stronger constraints on possible stellar companions were obtained by
\citet{van_eyken_ptf_2012} through high-resolution imaging with the
NIRC2 camera on the Keck II 10m telescope.  They reported 3-$\sigma$
$H$-band magnitude difference limits of 4.3, 6.4, and 8.9 at angular
separations of 0.25, 0.5, and 1.0 arcseconds (87, 175, and 349~AU).
They also detected a point source 7.0 magnitudes fainter than the
target, and 1.8$''$ to the north-east (which is not included in the
Gaia DR2 catalog). Due to its relative faintness, this object cannot
be the source of the shorter and longer-period TESS
signals.\footnote{This point source was claimed to be a potential
planetary-mass object \citep{schmidt_direct_2016}.  Subsequent
analysis of its colors showed that it is a background star
\citep{lee_evidence_2018}.}

\subsection{Photometric Binarity}

\begin{figure*}[t]
	\begin{center}
		\leavevmode
		\subfloat{
			\includegraphics[width=0.7\textwidth]{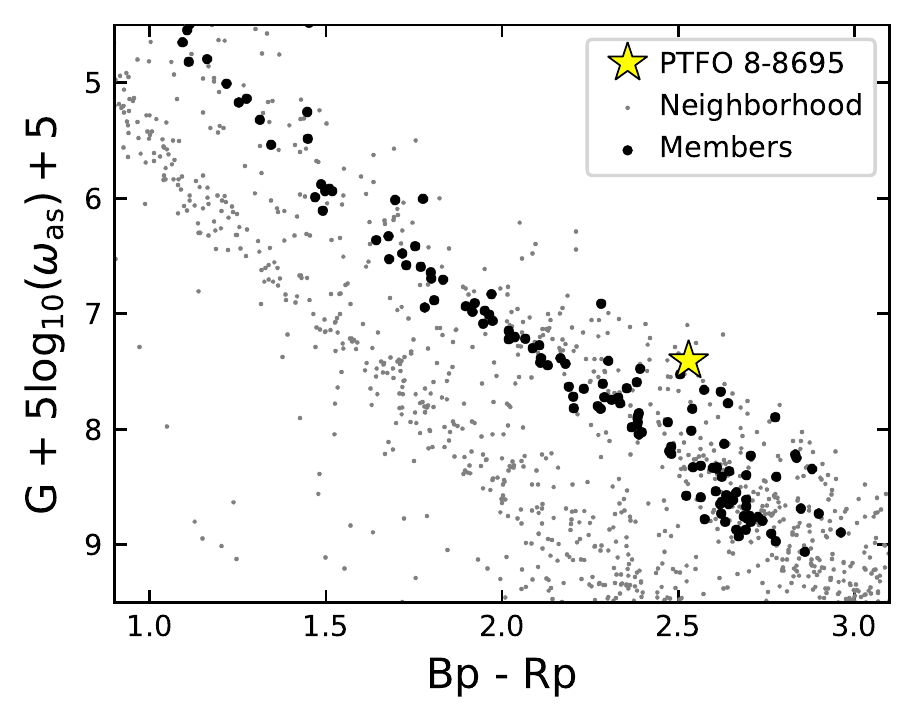}
		}
		
		\vspace{-0.7cm}
		\subfloat{
			\includegraphics[width=0.7\textwidth]{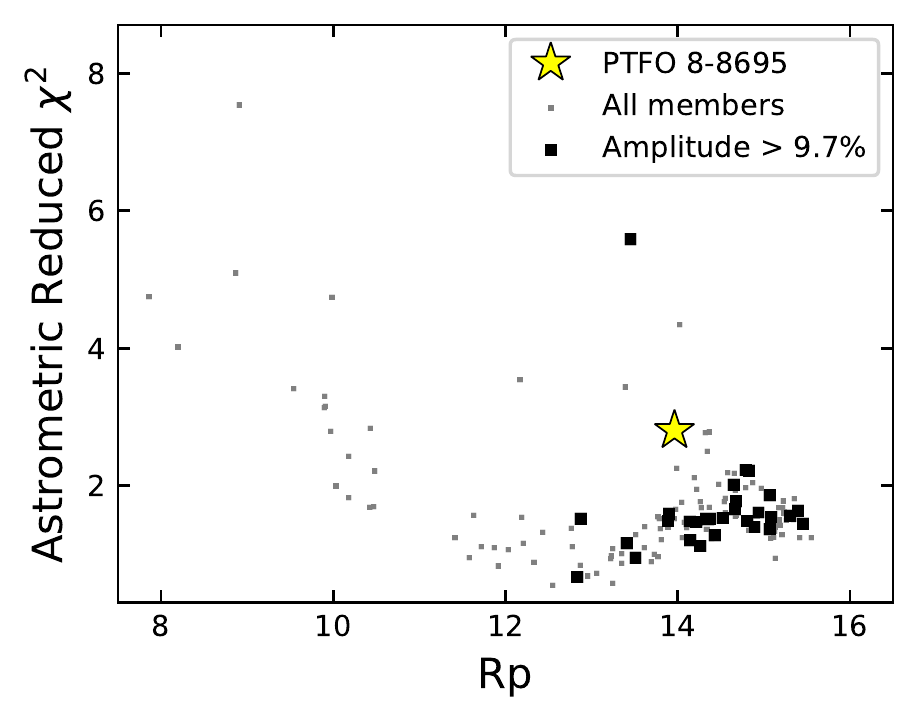}
		}
	\end{center}
	\vspace{-0.7cm}
	\caption{ {\bf Evidence for binarity in \ptfo}.
    {\it Top:} Hertzsprung-Russell diagram of \ptfo\ and late-type
    members of 25$\,$Ori-1. Black circles are members of the
    25$\,$Ori-1 group identified by \citet{kounkel_apogee2_2018}.
    Gray circles are stars in the ``neighborhood'', i.e., non-member
    stars for which the right ascension, declination, and parallax are
    within 5 standard deviations of the mean values for 25$\,$Ori-1.
    The neighborhood contains members of the Orion complex with its
    full spread of ages, in addition to field interlopers.  $G$
    denotes the Gaia broadband magnitude, $Bp$ Gaia blue, $Rp$ Gaia
    red, and $\omega_{\rm as}$ the parallax in arcseconds.  The
    $x$-axis limits are chosen to display only the K and M dwarfs,
    accentuating \ptfo's separation from the single-star sequence.
    {\it Bottom:} Astrometric goodness-of-fit versus $Rp$ magnitude
    for 25$\,$Ori-1 members.  The single-source astrometric model for
    \ptfo\ provides a poor fit to the data, which could be due to
    either stellar variability or binarity.  But since cluster members
    that are at least as variable as \ptfo\ show lower astrometric
    excesses (black squares), binarity is the likely reason.
		\label{fig:gaia}
	}
\end{figure*}

We also used the Gaia data to see if the observed luminosity of \ptfo\
is too high to be from a single star, i.e., if the object is a
``photometric binary.'' To assemble a set of stars coeval with \ptfo,
we used the 25$\,$Ori-1 members identified by
\citet{kounkel_apogee2_2018}, and discussed in
Section~\ref{subsec:hierarchical}.  To define a set of non-member
stars that nonetheless are subject to similar selection criteria, we
defined the reference ``neighborhood'' as the group of at most $10^4$
randomly selected non-member stars within 5 standard deviations of the
mean values of the right ascension, declination, and parallax of
25$\,$Ori-1.  We queried Gaia DR2 for these stars using
\texttt{astroquery} \citep{astroquery_2018}.  This yielded 1{,}819
neighbors.  While some of these stars may indeed be members of the
Orion complex, or even of 25$\,$Ori-1, enforcing this cut on positions
and parallaxes ensures that we are comparing stars with similar
amounts of interstellar reddening.

We examined the resulting five-dimensional distribution of right
ascension, declination, proper motion in both directions, and
parallax.  The first point we noted was that 25$\,$Ori-1 is a clearly
defined over-density in each dimension: the cluster was confirmed to
exist, and to be distinct from the neighborhood.  The second point we
noted is that \ptfo\ belongs to the cluster, based on its properties
in each of these dimensions.

Figure~\ref{fig:gaia} shows the HR diagram we constructed from the
data.  The diagram shows that \ptfo\ is $\approx$0.75 magnitudes
brighter than the average 25$\,$Ori-1 star of the same color.  In
other words, it is about twice as bright as expected for a single star
in the cluster.  It also seems to be part of a ``photometric binary''
track that runs above and parallel to the main track.

The implication is that either {\it (i)} \ptfo\ is notably younger
than the kinematically identical 25$\,$Ori-1 members, or {\it (ii)}
\ptfo\ is a binary with two components of nearly equal brightness.
Since there is no other reason to suspect an age difference, and
because the source showed two separate photometric signals with
similar but distinct periods, the binary interpretation seems more
probable.

\subsection{Astrometric Binarity}

A separate line of evidence for binarity is the Gaia DR2 astrometry.
As noted in Section~\ref{sec:observations}, the Gaia DR2 astrometric
solution for \ptfo\ shows a 10.3$\sigma$ ``astrometric excess'', a
parameter that quantifies the degree to which a single-star model
fails to fit the astrometric measurements.  Specifically, the
single-source astrometric model yielded $\chi^2=325.2$.  There are 121
astrometric measurements, and 5 free parameters, and therefore 116
degrees of freedom. The reduced $\chi^2$ is 2.80.  The majority of
stars with comparable brightness in Gaia do not show such poor
goodness-of-fit \citep[see][Appendix A]{lindegren_gaiasoln_2018}.

Potential explanations for the poor astrometric fit include
photometric variability and unresolved stellar binarity \citep[{\it
e.g.},][]{rizzuto_ZEIT8_2018,belokurov_unresolved_2020}.  If
photometric variability were the cause, we would expect stars of
similar brightness in the same kinematic group of Orion to show
similar astrometric excesses, because the majority of young stars are
highly variable.

Using the same 149 members in the 25$\,$Ori-1 subgroup, we calculated
the astrometric reduced $\chi^2$ for each member.  We then queried the
CDIPS light curve database at MAST \citep{bouma_cluster_2019} to find
the subset of members that were at least as variable as \ptfo.  We
measured the variability amplitude by taking the difference between
the $95^{\rm th}$ and $5^{\rm th}$ percentiles of the flux
measurements.  This yielded 30 stars of equal or greater variability.
The lower panel of Figure~\ref{fig:gaia} shows the reduced $\chi^2$ as
a function of stellar brightness.  \ptfo\ is in the upper 90$^{\rm
th}$ percentile of stars showing astrometric excesses within the
25$\,$Ori-1 group.  Relative to other M-dwarf group members with
comparable brightnesses and variability characteristics, \ptfo\ still
stands out by virtue of its failure to conform to a single-star
astrometric model. This supports the interpretation that \ptfo\ is a
binary star.

Performing the same analysis using the renormalized unit weight error
(RUWE\footnote{ See the Gaia DPAC technical note
GAIA-C3-TN-LU-LL-124-01,
\url{http://www.rssd.esa.int/doc_fetch.php?id=3757412}, accessed
2020-04-27. }) rather than the reduced $\chi^2$ yielded similar
results.  \ptfo\ has a RUWE of 1{.}22, which corresponds to the
93$^{\rm rd}$ percentile of 25$\,$Ori-1 members.  Two of thirty stars
with variability amplitudes greater than 9.7\% showed higher RUWE.
One was CVSO~35, which has a TESS light curve that varies by 2
magnitudes.  CVSO~35 also shows a strong Wide-field Infrared Survey
Exoplorer (WISE, \citealt{wright_WISE_2010}) IR excess as well as a
10$\mu$m silicate emission feature \citep{mauco_herschel_2018}.  The
other is GAIA DR2 3222210363837122048.

We will have to wait for the next data release of the Gaia mission for
a more definitive determination of whether the astrometric excess is
caused by stellar binarity or photometric variability.  Nonetheless
the fact that comparably variable stars do not show comparably large
astrometric excesses suggests that stellar binarity is indeed the root
cause.

\subsection{Radial Velocity Binarity}

Radial velocity (RV) measurements over sufficiently long timescales
could also reveal the presence of multiple stars in this system.
Unfortunately, the available RV data for \ptfo\ is sparse, presumably
due to the difficulties of performing RV observations of such a faint
and rapidly rotating star.  The RV datasets with the longest time
baselines we could find in the literature were those reported by
\citet{van_eyken_ptf_2012}.  These included 5 Keck/HIRES measurements
acquired over 10 days in April 2011, and 4 HET/HRS measurements
acquired over 10 days in February 2011.  The root-mean-squared RV over
each 10-day span was $\approx$$2\,{\rm km}\,{\rm s}^{-1}$, consistent
with the measurement precision.  Although \citet{van_eyken_ptf_2012}
tried a CCF-based RV reduction technique, they eventually found that
manually selecting absorption lines and measuring line centroids was
more effective.  While \citet{yu_tests_2015} acquired 22 further
Keck/HIRES spectra over one night in December 2013, those points were
not reduced to velocities. Further radial-velocity observations could
potentially confirm or refute the presence of binary companions.

\section{Discussion}
\label{sec:discussion}

\subsection{Longer-Period Signal}

The standard interpretation for 11.98$\,$hr nearly sinusoidal
modulations of a pre-main-sequence M dwarf is stellar rotation.  This
is the dominant signal in the system with 10\% amplitude, and there is
no evidence to suggest that this signal has any other origin.

In their report on the discovery of the unusual photometric
variability, \citet{van_eyken_ptf_2012} saw an alias of the
longer-period signal ({\it e.g.}, their Figure~7), in the form of a
peak in the periodogram at $0.9985 \pm 0.0061\,$days. They ascribed it
to their observing cadence, because of its close correspondence to the
sidereal day.  Our pixel-level analysis showed that the signal is
specific to only pixels near \ptfo, and no other pixels.  We therefore
conclude that the signal is not an artifact of systematic errors.

We are not the first to reach the conclusion that the long period
sinusoidal modulation is astrophysical.  \citet{koen_multicolour_2015}
identified the same modes and aliases as \citet{van_eyken_ptf_2012},
and argued that the signal was astrophysical, even if the exact period
was still unclear.  Using photometry from the YETI global telescope
network, \citet{raetz_yeti_2016} came to the conclusion that the
$0.50\,{\rm d}$ signal was indeed from stellar rotation.  The TESS
data strongly support this conclusion.

\subsection{Shorter-Period Signal, Including the ``Dip''}

\begin{figure}[t]
	\begin{center}
		\leavevmode
		\includegraphics[width=0.48\textwidth]{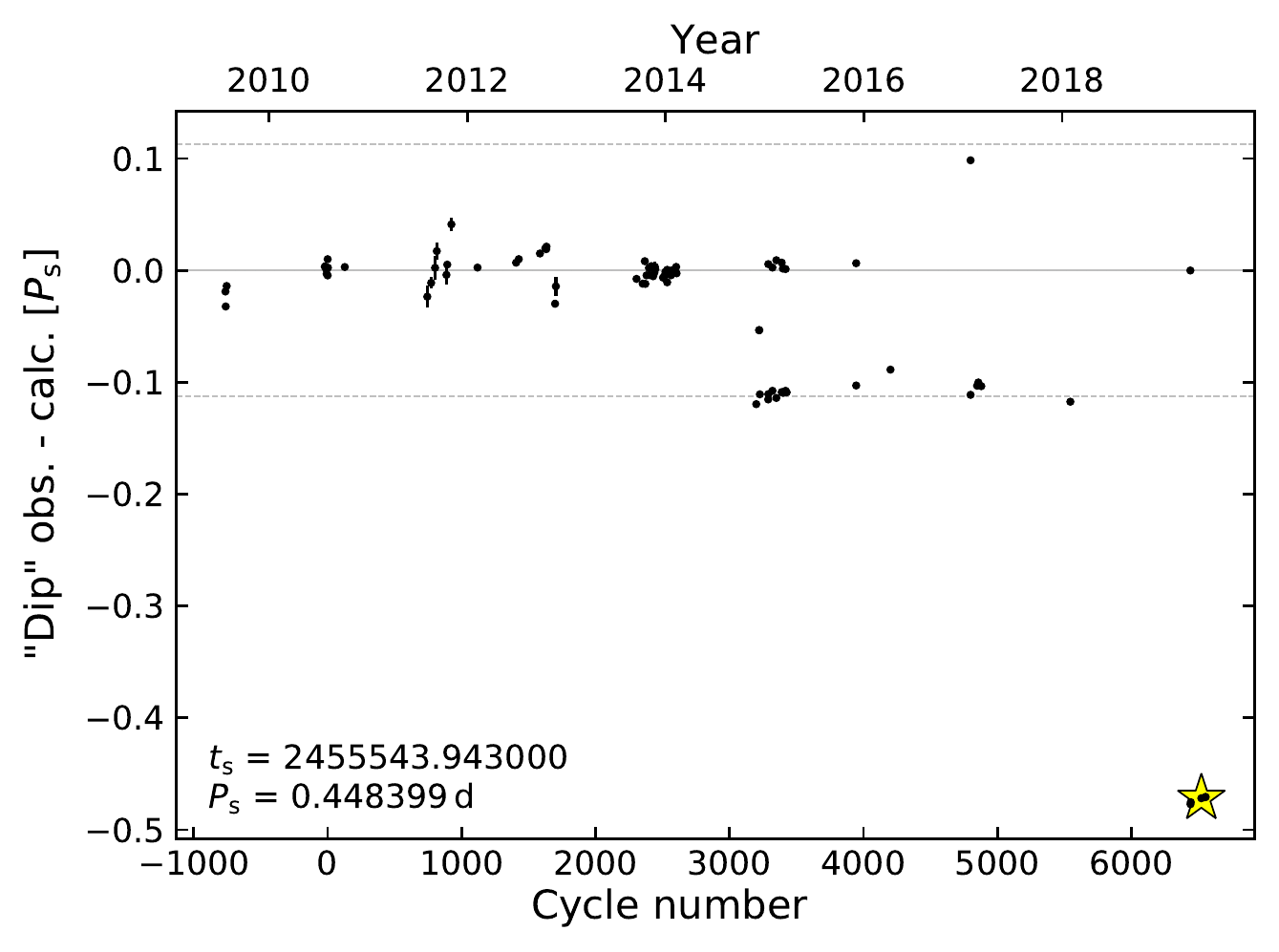}
	\end{center}
	\vspace{-0.7cm}
	\caption{
    {\bf Timing residuals for \ptfob\ based on a decade of
    monitoring.} Black points are times of dips, minus the indicated
    linear ephemeris.  The phase of the shorter-period signal is
    plotted on the $y$-axis. The star symbol represents the TESS
    ephemeris.  Dips were observed by \citet{van_eyken_ptf_2012},
    \citet{ciardi_followup_2015}, \citet{yu_tests_2015},
    \citet{raetz_yeti_2016}, \citet{onitsuka_multicolor_2017}, and
    \citet{tanimoto_evidence_2020}.  Certain dips ({\it e.g.}, the one
    at phase 0 in mid-2019) are consistent with noise, and were likely
    reported because dips were expected, rather than convincingly
    observed.  Horizontal dashed lines are drawn at $\pm (P_{\rm \ell}
    - P_{\rm s})/P_{\rm s}$, highlighting a possible numerical
    coincidence.  The orbital phase observed by TESS (lower-right) is
    consistent with that of \citet{tanimoto_evidence_2020}.
		\label{fig:o_minus_c}
	}
\end{figure}

The TESS light curve shows a dip that lasts about 45 minutes, and
recurs every 10.76 hours
(Figures~\ref{fig:splitsignal},~\ref{fig:splitsignalii},~\ref{fig:phasefold}).
The dip duration is roughly the same as that observed by previous
investigators \citep{van_eyken_ptf_2012,yu_tests_2015}.  The 1.2\%
depth is similar to what has been observed in the near-infrared
\citep{onitsuka_multicolor_2017}.  However the dip depth seems likely
to have evolved over time between being not present at all, to a
maximum of $\approx$5\% \citep[{\it
e.g.},][]{koen_multicolour_2015,yu_tests_2015,tanimoto_evidence_2020}.

An interesting feature of the sequence of dips is that the phase of
the dips has been observed to change with time \citep{yu_tests_2015}.
In fact, \citet{tanimoto_evidence_2020} provided stark evidence for
different behavior altogether: over a timespan of years, the dip
``split'' into distinct groups at particular phases.  See, for
instance, their Figures~2 through~4.  Fitting a decade of
observations, they provided the following constant-period ephemeris,
which we did not find any need to update:
\begin{align}
t_0\ {\rm BJD}_{\rm TDB} &= 2455543.943 \pm 0.002 \nonumber \\
P &= 0.4483993 \pm 0.0000006\,{\rm d}.
\label{eq:ephem}
\end{align}

Figure~\ref{fig:o_minus_c} shows the differences bewteen the observed
``mid-transit'' times of the dips and the times calculated using
Equation~\ref{eq:ephem}.  The phase of the dips seen by TESS (yellow
star) agrees with the independent December 2018 measurements by
\citet{tanimoto_evidence_2020}: either the dip abruptly shifted phase
over the past decade or, more likely, there are multiple dips that
have come and gone at different phases.

Figure~\ref{fig:o_minus_c} shows two additional strange features: {\it
(i)}  multiple dips per cycle, and {\it (ii)} a set of dips at a phase
that is numerically coincident with $(P_{\rm \ell} - P_{\rm s}) /
P_{\rm s}$.  The observation of multiple dips per cycle in 2015 was
seen independently by both \citet{yu_tests_2015} and
\citet{tanimoto_evidence_2020}.  It therefore seems credible.
Inspecting the \citet{tanimoto_evidence_2020} light curves, the claim
of multiple dips per cycle in December 2018 at phase 0 and $-0{.}47$
seems less plausible. The dips at phase $\text{-}0{.}47$ are strongly
detected, while the suggested dip at phase 0 is not clearly deteceted.

We are not sure what to make of the numerical coincidence.  The ratio
of long to short periods is roughly 10:9.  It is not clear that this
would obviously translate into an observational bias unless, by some
fluke, three season's worth of observations managed to only observe
every ninth dip.  This is of course not the case, and we therefore
leave this curiosity as observation {\it sans} interpretation.

\subsection{Short Period Modulation Outside of Dips}
Visually, the out-of-dip modulation at the 10.76$\,$hr period
resembles a slightly asymmetric sinusoid (Figure~\ref{fig:phasefold}).
The best model has non-zero amplitudes for both the first and second
harmonics (Table~2).  The third harmonic is formally present with
marginal ($\approx$2$\sigma$) significance.  The first sine and cosine
harmonic both have amplitudes of roughly $0.90\pm0.04\%$.  The second
sine harmonic has amplitude $0.16 \pm 0.04\%$, so is non-zero at a
significance of $\approx$4$\sigma$.  The second cosine harmonic has an
amplitude of $-0.55 \pm 0.03\%$.  In our sign convention, the fact
that it is negative means that this component peaks at phase 0.25 and
0.75, i.e., the quadratures of the orbit.

\subsubsection{Ellipsoidal Variability?}
If there were a giant planet transiting \ptfo, it would tidally
distort the host star, and cause ellipsoidal photometric modulations
that peak at the quadratures \citep[see][]{shporer_astrophysics_2017}.
Interpreting the second cosine harmonic as planet-induced tidal
distortion, it would imply a minimum planet mass $M_{\rm p} \sin i$ of
$3.8\,M_{\rm Jup}$.  For this estimate, we assumed $R_\star = 1.39
R_\odot$, and $M_\star = 0.39 M_\odot$ \citep{van_eyken_ptf_2012}.
This ellipsoidal amplitude is larger than the typical modulations
induced by close-in giant planets because the host star is puffy, and
still on the pre-main-sequence.

The planetary interpretation however does not readily explain the
large first sine and cosine harmonics.  Interpreting the sine
component as Doppler beaming would imply a secondary mass greater than
the primary ($0.86\,M_\odot$).  Interpreting the cosine component as
reflected or emitted light from the planet's surface is nonsensical
because the sign is wrong---the planet would need to be {\it
absorbing} light.

\subsubsection{Similar Light Curves}
\label{subsec:dipstars}

\begin{figure*}[hbtp]
	\begin{center}
		\leavevmode
		\includegraphics[width=1\textwidth]{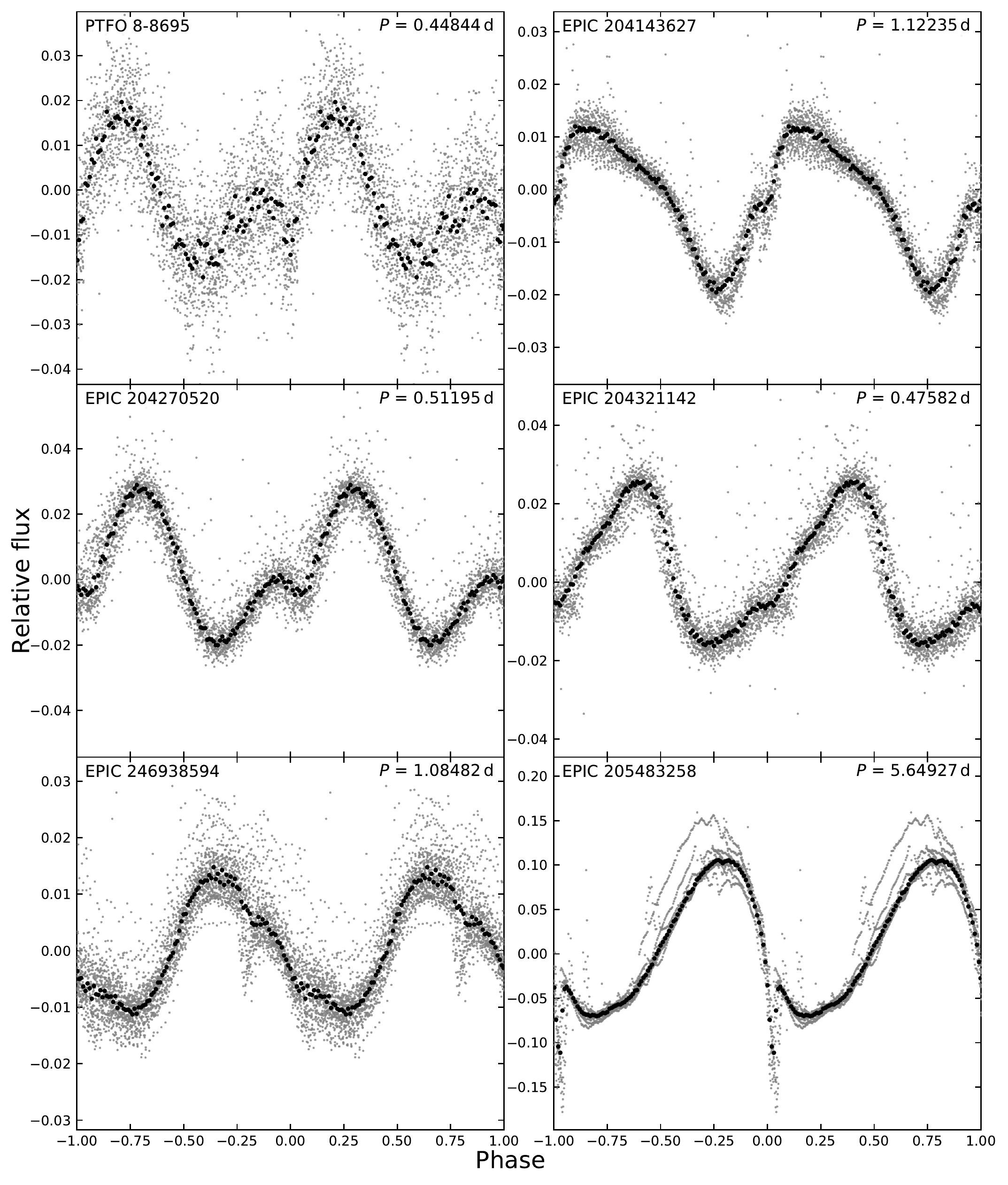}
	\end{center}
	\vspace{-0.7cm}
  \caption{ {\bf \ptfo\ and its brethren.}
    Shown are the light curves of five transient and persistent flux
    dip stars, selected based on their similarity to the short-period
    signal of \ptfo.  The stars are EPIC 204143627, EPIC 204270520,
    EPIC 204321142, EPIC 246938594, and EPIC 205483258 (RIK-210).
    RIK-210 has the longest period of any of these objects.  All the
    analogs displayed are either in Taurus or Upper~Sco, and meet the
    characteristics of Section~\ref{subsec:dipstars}.  These objects
    were originally reported by \citet{stauffer_orbiting_2017},
    \citet{david_transient_2017}, and \citet{rebull_usco_2018}.
		\label{fig:brethren}
	}
\end{figure*}

When physical explanations are not forthcoming, we often resort to
taxonomy.  By searching the literature, we have found about a dozen
light curves with similar morphologies to \ptfo, drawn from surveys of
low-mass weak-lined T Tauri stars in regions including $\rho$ Oph,
Upper~Sco, Taurus, and perhaps the Pleiades
\citep{rebull_rotation_2016,david_transient_2017,stauffer_orbiting_2017,stauffer_rotevol_2018,rebull_usco_2018,rebull_rotation_2020}.
These surveys were performed using K2 \citep{howell_k2_2014}.  We
downloaded some of these light curves from MAST, opting for the
EVEREST reductions \citep{luger_everest_2016,luger_update_2018}.  They
are plotted in Figure~\ref{fig:brethren}.

These light curves have been phenomenologically classified as
``persistent flux dips'' or ``transient flux dips'', based on whether
their depths and durations show variability over the 90-day K2
campaigns \citep{stauffer_orbiting_2017}.  In the terminology of
\citet{stauffer_orbiting_2017}, these objects are morphologically
distinct from ``scallop shell'' light curves, and are present in stars
at more advanced evolutionary disk stages than the ``dipper'' stars
\citep{ansdell_young_2016,cody_manyfaceted_2018}.  The persistent and
transient flux dip stars all show angular dips that are cannot be
explained as the effects of starspots.  These stars typically have the
following things in common:
\begin{enumerate}[topsep=0.5ex,itemsep=-0.5ex,partopsep=1ex,parsep=1.5ex]
  \item They are weak-lined T Tauri stars.
  \item The spectral type is M2 to M5 ({\it e.g.},
    \citealt{rebull_usco_2018},~Figure~20).
  \item The age is typically $\lesssim$ 100$\,$Myr.\footnote{ At
    present, the oldest observed ``scallops'' are in the Pleaides
    \citep{rebull_rotation_2016}. One of these, EPIC 211013604, might
    meet the ``persistent dip'' classification.  If so, it is also the
    oldest known.} 
  \item The light curves show shallow, angular dips, usually
    superposed on large-amplitude smooth variability. The latter is
    interpreted as stellar rotation.
  \item The rotation is rapid, with a period that is usually between
    0.5 and 2.0 days.
  \item There is rarely any infrared excess that is detectable in the
    WISE data (never any W4 detection; only a few W3 detections).
  \item They sometimes show multiple dips per cycle.
  \item The dip depths, durations, and phases can vary over just a few
    cycles ({\it e.g.}, EPIC 204143627).
  \item The dip depths can change after flares.
  \item They are rare at a population level, with an occurrence rate
    of $\lesssim 1\%$ of young M2-M5 stars \citep{rebull_usco_2018}.
\end{enumerate}
The 10.76$\,$hr signal in \ptfo\ meets all of these criteria.  This is
the first connection of \ptfo\ with this class of objects, likely
because the TESS data enabled us to put the dips in the context of the
asymmetric out-of-dip modulation.

There are two crucial additional points concerning the transient flux
dips.  First, the dip durations seem to scale linearly with the
photometric periods \citep[][Figure~26]{stauffer_orbiting_2017}.  In
contrast, the transit duration $T$ of a small obstructing object
across the stellar disk scales as $T \propto R_\star
(P/M_\star)^{1/3}$ \citep{winn_exoplanet_2010}.  While the shortest
period $\approx$0.5-day transient flux dip stars have dip durations
consistent with point sources, at longer periods of 1 to 5 days the
dip durations become many hours, which is too long to be caused by
planetary transits.

Second, approximately 40-50\% of the transient flux dip stars
discovered in $\rho$~Oph and Upper~Sco show two Lomb-Scargle periods,
and so are apparently binaries
\citep[][Table~1]{stauffer_orbiting_2017}.  This is higher than the
main-sequence companion fraction of ${\rm
CF}_{0.1\text{-}0.5\,M_\odot}^{\rm MS} = 33 \pm 5\%$
\citep{henry_solar_2006,duchene_stellar_2013,winters_solar_2019}.
Low-mass pre-main-sequence stars however have been shown to companion
fractions up to twice as high in dispersed clusters such as Upper~Sco
and Taurus \citep{kraus_mapping_2008,kraus_mapping_2011}.  A
high-resolution imaging survey would be interesting, to determine
whether the transient flux dip stars truly have distinct
population-level binarity properties relative to other young low-mass
stars.

\subsection{Physical Interpretation}
\label{subsec:physical}

The evidence for binarity in \ptfo\ is as follows.  First, the star is
twice as bright as stars of the same color in its kinematic group
(Figure~\ref{fig:gaia}).  Second, it shows two distinct photometric
signals.  These points alone suggest binarity
\citep{stauffer_rotevol_2018}.  For the case of \ptfo, there is a
third line of evidence: the Gaia DR2 entry for \ptfo\ reports a poor
fit of the single-star model to the astrometric data.  While this
could be caused by stellar variability, other cluster members that are
just as variable do not typically show the same level of excess
astrometric motion.  Therefore the astrometric excess is a suggestive
third line of evidence for binarity in \ptfo. To us, the evidence
leads to the conclusion that \ptfo\ is a nearly equal-mass binary
consisting of two rapidly rotating stars.

Based on the lack of an infrared excess seen by \citet{yu_tests_2015},
the primordial gas disks of both stars in \ptfo\ seem to be have been
depleted.  This is consistent with the $8.5\pm1.2\,{\rm Myr}$ age of
the 25$\,$Ori-1 group, and the rapid rate at which stars show lose
their disks between 1 and 10$\,$Myr \citep[{\it
e.g.},][]{hernandez_spitzer_ob1_2007}.  The stars are therefore
presumably no longer magnetically locked to their disks.  This is also
suggested by the $\approx$half-day periodicities of both rotation
signals: young disked M dwarfs typically rotate with periods of two
days or more due to magnetic locking \citep[{\it
e.g.},][]{rebull_rotation_2020}.  If the two stars are within
$\approx$50$\,$AU of each other, as required by the NIRC2 adaptive
optics imaging, then it would also be expected that the stars would
have truncated the outer edges of their respective disks, in a manner
seen at the population level in exoplanetary systems
\citep{kraus_impact_2016,moe_impact_2019}.

The key question is what causes the transient dips. This is
an unsolved problem not only for \ptfo\ but also for the emerging
class of similar young and rapidly rotating M-dwarfs.  Many possible
explanations were discussed by \citet{rebull_rotation_2016},
\citet{david_transient_2017}, \citet{stauffer_orbiting_2017}, and
\citet{zhan_complex_2019}.  Among the {\it disfavored} explanations
are that the dips are caused by
{\it (i)} eclipsing binaries;
{\it (ii)} ``dipper''-flavor Class-I or Class-II disks;
{\it (iii)} eclipses of prominences;
{\it (iv)} high-latitude accretion hotspots;
{\it (v)} high-latitude starspots;
or
{\it (vi)} dust clouds of plausible composition.
We also view the possibility of {\it (vii)} tidally disrupted
planetary or cometary material to be implausible, given the
synchronicity between dip and rotation periods seen across many
systems.

The explanations that are not yet ruled out include {\it (i)}
transiting clumps of gas at the Keplerian corotation radius; {\it
(ii)} transits of enshrouded protoplanets; {\it (iii)} occultations of
starspots by an optically thick disk.  The first and last explanations
have added appeal because they are flexible enough to explain not only
the transient and persistent-dip M-dwarfs, but also the ``scallop
shell'' M-dwarfs \citep{stauffer_orbiting_2017}.  Despite this appeal,
the possibility of distinct mechanisms explaining these distinct
variability classes remains open.

The evolution of \ptfo\ over the past decade
(Figure~\ref{fig:o_minus_c}) could offer important hints.
Specifically, \ptfo's transition between having none, one, and
multiple dips per cycle seems important.  It strains the ``enshrouded
protoplanet'' interpretation, because there are no known processes
that cause a planet's orbital phase to jump.  The dips would then need
to be caused by material that was somehow disrupted from the planet,
but somehow remained co-orbital for an extended duration. This seems
implausible.

\section{Conclusions}
\label{sec:conclusions}

The combination of TESS and Gaia data has clarified a few things about
the \ptfo\ system.  Our main results are as follows.
\begin{itemize}
  \item {\it The TESS light curve shows two periodic signals.} The
    ``long'' signal is a 10\% peak-to-peak sinusoid that repeats every
    11.98$\,$hr.  The ``short'' signal is a 4\% peak-to-peak ``dip +
    asymmetric sinusoid'' that repeats every 10.76$\,$hr. The signals
    beat, and therefore cannot be an artifact linked to data
    processing.  Within the angular resolution of the Gaia source
    catalog, both signals originate from \ptfo.
  \item {\it The Gaia data imply binarity.} Relative to stars in its
    kinematic group, \ptfo\ is a photometric binary
    (Figure~\ref{fig:gaia}, top).  Relative to stars in its group that
    are at least as photometrically variable, \ptfo\ also shows signs
    of astrometric binarity (Figure~\ref{fig:gaia}, bottom).
  \item {\it The orbital phase of the dip has changed since the
    discovery by \citet{van_eyken_ptf_2012}.} As shown in
    Figure~\ref{fig:o_minus_c}, the phase seems to have jumped,
    perhaps twice. This agrees with the recent study by
    \citet{tanimoto_evidence_2020}.
  \item {\it All properties of \ptfo\ are consistent with the emerging
    class of transient and persistent flux dip stars.} Analogous
    light curves are shown in Figure~\ref{fig:brethren}.  Properties of
    this variability class are enumerated in
    Section~\ref{subsec:dipstars}.
\end{itemize}

The physical mechanism that explains the transient and persistent flux
dips is unresolved. Our preferred explanations include transiting
clumps of gas at the Keplerian corotation radius, and occultations of
starspots by a tenuous gas disk \citep[{\it
e.g.},][]{stauffer_orbiting_2017,david_transient_2017,zhan_complex_2019}.
The jumping orbital phase disfavors the explanation of an enshrouded,
transiting protoplanet.  Though \ptfob\ may not be a planet, as we and
others had hoped, understanding \ptfo\ and its analogs is a worthy
problem.  It might even teach us about the birth environments of the
majority of habitable-zone Earth-sized planets
\citep{dressing_occurrence_2013}.


\acknowledgements
When this manuscript was at an advanced stage, we received notice of a
paper by \citet{koen_2020} that was in press at the {\it Monthly
Notices} before submission of our manuscript.  Our studies
independently reached the same conclusions: the TESS light curve shows
two periodic signals, and the properties of \ptfo\ are consistent with
the emerging class of transient and persistent flux dip stars.
\citet{koen_2020} reached these conclusions by modeling the TESS light
curve as a truncated sum of Fourier terms, and concluded that the two
signals are most simply interpreted as coming from two stars.  Our
analysis of the Gaia data provides independent support for the
conclusion that \ptfo\ is a binary. We also note the agreement between
the TESS dip ephemeris and that from \citet{tanimoto_evidence_2020}.
\\
\\
%
%
%
The authors thank D.~Fabrycky, S.~Mahadevan, G.~Stef\'ansson, and
A.~Vanderburg for helpful calculations, observations, and suggestions.
We also thank the Heising-Simons Foundation for
their generous support of this work.
\ptfo\ was included on the TESS ``short-cadence'' target list thanks
to the Guest Investigator programs of S.\ Czesla and C.\ Huang
(G011128 and G011132 respectively).
Resources supporting this work were provided by the NASA High-End Computing (HEC) Program through the NASA Advanced Supercomputing (NAS) Division at Ames Research Center for the production of the SPOC data products.
The Digitized Sky Survey was produced at the Space Telescope Science
Institute under U.S. Government grant NAG W-2166.
Figure~\ref{fig:scene} is based on photographic data obtained using
the Oschin Schmidt Telescope on Palomar Mountain.

\software{
  \texttt{astrobase} \citep{bhatti_astrobase_2018},
  \texttt{astropy} \citep{astropy_2018},
  \texttt{astroquery} \citep{astroquery_2018},
  \texttt{cdips-pipeline} \citep{bhatti_cdips-pipeline_2019},
  \texttt{corner} \citep{corner_2016},
  \texttt{exoplanet} \citep{exoplanet:agol19},
  \texttt{exoplanet} \citep{exoplanet:exoplanet}, and its
  dependencies \citep{exoplanet:agol19, exoplanet:kipping13, exoplanet:luger18,
  	exoplanet:theano},
  \texttt{IPython} \citep{perez_2007},
	\texttt{lightkurve} \citep{lightkurve_2018},
  \texttt{matplotlib} \citep{hunter_matplotlib_2007}, 
  \texttt{MESA} \citep{paxton_modules_2011,paxton_modules_2013,paxton_modules_2015},
  \texttt{numpy} \citep{walt_numpy_2011}, 
  \texttt{pandas} \citep{mckinney-proc-scipy-2010},
  \texttt{pyGAM} \citep{serven_pygam_2018_1476122},
  \texttt{PyMC3} \citep{salvatier_2016_PyMC3},
  \texttt{scipy} \citep{jones_scipy_2001},
  \texttt{SPOC R4.0} \citep{jenkins_tess_2016},
  \texttt{tesscut} \citep{brasseur_astrocut_2019},
  \texttt{wotan} \citep{hippke_wotan_2019}.
}

\facilities{
 	{\it Astrometry}:
 	Gaia \citep{gaia_collaboration_gaia_2016,gaia_collaboration_gaia_2018}.
 	{\it Imaging}:
  Second Generation Digitized Sky Survey,
 	Keck:II~(NIRC2; \url{www2.keck.hawaii.edu/inst/nirc2}).
 	{\it Spectroscopy}:
 	Keck:I~(HIRES; \citealt{vogt_hires_1994}).
 	{\it Photometry}:
 	TESS \citep{ricker_transiting_2015}.
}

\startlongtable
\begin{deluxetable*}{lrrrrrrrr}
%
%
\tablenum{1}
\tablecaption{Model Comparison.}
\label{tab:modelcompare}
\tablehead{
\colhead{Description} &
\colhead{$N_{\rm s}$} &
\colhead{$N_{\rm \ell}$} &
\colhead{$N_{\rm data}$} &
\colhead{$N_{\rm param}$} &
\colhead{$\chi^2$} &
\colhead{$\chi_{\rm red}^2$} &
\colhead{BIC} &
\colhead{$\Delta$BIC}
}
%
\startdata
Favored    & 3 &  2 &   2585 &      21 &  3102.4 &     1.210 &  3267.4 &     0.0 \\
\hline
Disfavored  &  2 &  3 &   2585 &      21 &  3179.0 &     1.240 &  3344.0 &    76.6 \\
---         &  2 &  2 &   2585 &      19 &  3237.4 &     1.262 &  3386.7 &   119.3 \\
---         &  3 &  3 &   2585 &      23 &  3217.1 &     1.256 &  3397.9 &   130.4 \\
---         &  2 &  1 &   2585 &      17 &  3312.6 &     1.290 &  3446.1 &   178.7 \\
---         &  3 &  1 &   2585 &      19 &  3397.5 &     1.324 &  3546.8 &   279.4 \\
---         &  1 &  2 &   2585 &      17 &  4101.2 &     1.597 &  4234.8 &   967.3 \\
---         &  1 &  3 &   2585 &      19 &  4160.8 &     1.622 &  4310.1 &  1042.7 \\
---         &  1 &  1 &   2585 &      15 &  4318.4 &     1.680 &  4436.2 &  1168.8 \\
\enddata
\tablecomments{
	$N_{\rm s}$ and $N_{\rm \ell}$ are the number of harmonics at the short and long periods, respectively.
	$N_{\rm data}$ is the number of fitted flux measurements.
	$N_{\rm param}$ is the number of free parameters in the model.
	The Bayesian information criterion (BIC) and the difference from the maximum $\Delta {\rm BIC}$ are also listed.
}
\vspace{-1cm}
\end{deluxetable*}

\startlongtable
\begin{deluxetable*}{lllrrrr}
\tablecaption{ Best-fit model priors and posteriors. }
\label{tab:posterior}
%
%
\tablenum{2}
\tablehead{
  \colhead{Param.} & 
  \colhead{Unit} & 
  \colhead{Prior} & 
  \colhead{Mean} & 
  \colhead{Std{.} Dev.} &
  \colhead{3$^{\rm rd}$ Pct.} &
  \colhead{97$^{\rm th}$ Pct.}
}
\startdata
{\it Sampled} & & & & & & \\
\hline
$P_{\rm s}$ & d & $\mathcal{N}(0.4485; 0.0010)$ & 0.4484613 & 0.0000460 & 0.4483731 & 0.4485416 \\
$t_{\rm s}^{(1)}$ & d & $\mathcal{N}(0.438096; 0.0020)$ & 0.4388368 & 0.0011286 & 0.4367929 & 0.4410297 \\
$R_{\rm p}/R_\star$ & -- & $\mathcal{N}(0.1100; 0.0110)$ & 0.11171 & 0.00679 & 0.09950 & 0.12437 \\
$b$ & -- & $\mathcal{U}(0; 1+R_{\mathrm{p}}/R_\star)$ & 0.8205 & 0.0523 & 0.7188 & 0.9071 \\
$u_1$ & -- & (2) & 0.693 & 0.501 & 0. & 1.638 \\
$u_2$ & -- & (2) & -0.01 & 0.429 & -0.804 & 0.806 \\
Mean & -- & $\mathcal{U}(-0.01; 0.01)$ & -0.001019 & 0.000185 & -0.001365 & -0.000669 \\
$R_\star$ & $R_\odot$ & $\mathcal{T}(1.23; 0.40)$ & 1.20 & 0.40 & 0.44 & 1.90 \\
$M_\star$ & $M_\odot$ & $\mathcal{T}(0.39; 0.25)$ & 0.42 & 0.22 & 0. & 0.78 \\
$A_{\mathrm{s},0}$ & -- & $\mathcal{U}(-0.02; 0.02)$ & 0.009083 & 0.000371 & 0.008396 & 0.009763 \\
$B_{\mathrm{s},0}$ & -- & $\mathcal{U}(-0.02; 0.02)$ & 0.009696 & 0.000391 & 0.008914 & 0.010352 \\
$A_{\mathrm{s},1}$ & -- & $\mathcal{U}(-0.02; 0.02)$ & 0.001646 & 0.000351 & 0.000990 & 0.002297 \\
$B_{\mathrm{s},1}$ & -- & $\mathcal{U}(-0.02; 0.02)$ & -0.005456 & 0.000307 & -0.005998 & -0.004861 \\
$A_{\mathrm{s},2}$ & -- & $\mathcal{U}(-0.02; 0.02)$ & 0.000177 & 0.000252 & -0.000295 & 0.000655 \\
$B_{\mathrm{s},2}$ & -- & $\mathcal{U}(-0.02; 0.02)$ & -0.000581 & 0.000271 & -0.001110 & -0.0001 \\
$\phi_{\rm \ell}$ & rad & $\mathcal{U}(1.3721; 2.1575)$ & 1.80542 & 0.20468 & 1.47712 & 2.09634 \\
$\omega_{\rm \ell}$ & rad$\ $d$^{-1}$ & $\mathcal{N}(12.6054; 0.1261)$ & 12.588753 & 0.000972 & 12.586968 & 12.590517 \\
$A_{\mathrm{\ell},0}$ & -- & $\mathcal{U}(-0.06; 0.06)$ & 0.03929 & 0.004331 & 0.031501 & 0.045035 \\
$B_{\mathrm{\ell},0}$ & -- & $\mathcal{U}(-0.06; 0.06)$ & 0.019891 & 0.008161 & 0.0071 & 0.032232 \\
$A_{\mathrm{\ell},1}$ & -- & $\mathcal{U}(-0.02; 0.02)$ & 0.002189 & 0.000516 & 0.001203 & 0.003021 \\
$B_{\mathrm{\ell},1}$ & -- & $\mathcal{U}(-0.02; 0.02)$ & -0.002311 & 0.000496 & -0.003063 & -0.001364 \\
\hline
{\it Derived} & & & & & & \\
\hline
$\omega_{\rm s}$ & rad$\ $d$^{-1}$ & -- & 14.01054 & 0.00144 & 14.00803 & 14.01330 \\
$R_{\rm p}$ & $R_{\mathrm{Jup}}$ & -- & 1.30 & 0.44 & 0.53 & 2.16 \\
$a/R_\star$ & -- & -- & 1.81 & 3.17 & 0.35 & 3.29 \\
\enddata
\tablenotetext{}{
$\mathcal{U}$ denotes a uniform distribution,
$\mathcal{N}$ a normal distribution, and
$\mathcal{T}$ a truncated normal bounded between zero and an upper limit much larger than the mean.
  Note that $R_{\rm p}/R_\star$ has been corrected for the dilution by
  Star A and other neighboring stars, according to the PDCSAP
  lightcurve's \texttt{CROWDSAP} value (0{.}73) in the optimal aperture.
  (1) To convert mean TESS mid-transit time to ${\rm BJD}_{\rm TDB}$, add 2458468.2.
  (2) Quadratic limb-darkening prior from \citet{exoplanet:kipping13}, implemented by \citet{exoplanet:exoplanet}.
}
\vspace{0cm}
\end{deluxetable*}


\bibliographystyle{yahapj}                            
\bibliography{bibliography}

\listofchanges

\end{document}